# Generalized formal model of big data

N. Shakhovska[1], O. Veres[2], M. Hirnyak[2]

[1] *University of Economy, Bydgoszcz; e-mail: Nataliya.shakhovska@lpnu.ua*
[2] *Lviv Polytechnic National University; e-mail: Oleh.M.Veres@lpnu.ua*

*Abstract.* This article dwells on the basic characteristic features of the Big Data technologies. It is analyzed the existing definition of the "big data" term. The article proposes and describes the elements of the generalized formal model of big data. It is analyzed the peculiarities of the application of the proposed model components. It is described the fundamental differences between Big Data technology and business analytics.

Big Data is supported by the distributed file system Google File System technology, Cassandra, HBase, Lustre and ZFS, by the MapReduce and Hadoop programming constructs and many other solutions. According to the experts, such as McKinsey Institute, the manufacturing, healthcare, trade, administration and control of individual movements undergo the transformations under the influence of the Big Data.

## INTRODUCTION

These days there are a lot of Big Data sources. These are the data continuously received from the measuring devices, events from radio frequency identifiers, the message flow from social networks, meteorological data, Earth remote sensing data, data flows about the location of the cellular communication networks subscribers, audio and video recording devices. The mass distribution of the above mentioned technologies and innovative models using various kinds of devices and Internet services was the starting point for the Big Data penetration in almost all the spheres of human activity, primarily in research activities, the commercial sector and public administration.

Today the issues of the correct interpretation of the information flows are becoming more relevant and challenging at the same time. The volume of information is growing exponentially and its lion's share belongs to the unstructured data. The market of information technologies (IT) has reacted immediately. The large players acquired the most successful niche companies and began to develop tools for the Big Data. The number of startups exceeds all possible expectations [1].

Along with the growth of the computation power and the development of storage technologies the analysis of the Big Data is gradually becoming available to small and medium businesses and is no longer the exclusive prerogative of the large companies and research centers. To a large extent, the development of the cloud model computing contributes to it.

With further IT penetration into the business environment and the daily life, information flows (entitled to processing) continue to grow steadily. If today the Big Data are petabytes, tomorrow we will have to operate with exabytes, etc. In the observable future, it is obvious that the tools for such gigantic amounts of information still be too complex and expensive. It is considered that the modern software tools cannot handle such volumes within a reasonable time period. Obviously, the indicated range of values is very conditional and tends to increase upward as the computing hardware is continuously improving and becoming more accessible.

## THE ANALYSIS OF RECENT RESEARCHES AND PUBLICATIONS

The concept of the Big Data is not new, it is originated in the days of mainframes and the related scientific computing [1, 2]. As is well-known, knowledge-intensive computing has always been challenging. As a rule, it is inextricably linked to the processing of the large volumes of information.

However, directly the "big data" term has emerged relatively recently. It is among the few titles that has its quite reliable birthday on September 03, 2008. Then a special issue of the oldest British scientific journal Nature has been released. The journal is devoted to the search for the answer to the question: "How can the technologies affect to the scientific future that open up the opportunities to work with the big data?" A special issue summarized preliminary discussions about the data role of science in general and in e-science in particular [3].

According to the report by McKinsey Institute entitled "Big data: The next frontier for innovation, competition and productivity", the "big data" term refers to the datasets whose size exceeds the capacity of the conventional database (DB) for the extraction, storage, management and information analysis [1]. Global data repositories, of course, continue to grow. The presented report of an IDC analytical company entitled "Digital Universe Study" in the mid-2011 (which was sponsored by EMC company) assumes that the total global volume of the generated and replicated data may reach in 2011 about 1.8 zettabytes (1.8 trillion gigabytes). It is about 9 times more than what was established in 2006.

However, "big data" entails much more than just analyzing the huge volumes of information. The problem is not that organizations generate huge amounts of data, but that the most of them are presented in a format that does not fit well with the traditional structured database format. These are web-magazines, videos, text documents, machine code, or, for instance, mapping data. All of the above mentioned are available in many different stores, sometimes even outside the organization. As a result, the corporation may have access to a huge volume of its data and does not have the necessary tools to establish the relationships between these data and make them the basis of meaningful conclusions. In addition, the data is now updated more often and we have a situation that the traditional methods of information analysis cannot overtake the huge volumes of constantly updated data which ultimately paves the way for the Big Data technologies.

The concept of "big data" considers working with the huge information volume and diverse composition. This information is very frequently updated and located in different sources with the aim of the efficiency increasing, new product generation and competitiveness improving. A consulting company Forrester gives a brief definition: "Big Data combine techniques and technologies that extract sense of data at the extreme limits of practicality" [1].

Worldwide information volume is growing annually by at least 59%. The corporate data sets and all the technologies needed to create, store, transfer, analyze, archive and extract are considered "big data". This huge information volume expands the limits of storage and security servers, creating a huge challenge for the IT departments that need to be solved. Therefore, the data volume, variety and velocity are the important challenges in the management of Big Data (on which the business and IT leaders should focus).

eWEEK submits the definition proposed by the research company Gartner: "Big Data are characterized by volume, variety and pouring rate through networks of structured and unstructured data, into the processors and storage devices, as well as the data conversion to business consulting for the enterprises" [4]. These elements can be divided into three categories as defining characteristics of Big Data ("three V"): volume, variety and velocity.

Volume (terabytes, petabytes and eventually exabytes). An increasing amount of business data is a major shot to the IT systems that try to keep safe and accessible all of this information for further use. Increase of data volumes in enterprise systems causes a significant scope of transactions over the traditional and new types of data. A large data volume is the problem of storage and analysis.

Variety. The number of data types that need to be processed is increasing, namely, the possibility of simultaneous processing of different types of structured, semi-structured and unstructured data. The variety includes tabular data (database), hierarchical data, documents, emails, accounting records, video, static images, audio, financial transactions, and the like.

Velocity. This refers to the velocity at which the data are moved from the endpoints to the processing and storage means. Velocity means how fast is the data produced and how fast must data be processed to meet the demand, and how fast is data saved.

Analyst Dan Kusnetzky (Kusnetzky Group) argues that the "big data" term refers to the tools, processes and procedures allowing an organization to create, manage and control the very large data sets and repositories [4].

In addition to the volume, variety and velocity, there is another "v" which fits in the overall data picture. It is a value. Values provide an accurate analysis of the large volumes of data. They help the businessmen to make the right decision at the appropriate time.

The "big data" term describes the data sets with a possible exponential growth that are too large, too plain or too unstructured for the analysis by traditional methods [6]. Big Data in information technology is a set of processing methods and means of structured and unstructured, dynamic, heterogeneous Big Data for their analysis and use of the decision support [7].

Bill Inmon considers the concept of "big data" as a new information technology [8]. Big Data is a technology that has the following features:
- it is possible to process a very large volume of data,
- the data media are inexpensive,
- data are managed by the "Roman Census" method,
- data managed with the Big Data are unstructured.

It is hard to find the industry to which the problem of Big Data would be irrelevant. The ability to handle the large volumes of information, analyze the relationships between them and to make informed decisions, on the one hand, carries the potential for the companies from different verticals to increase the attractiveness and profitability, efficiency. On the other hand, this is a great opportunity for the additional income to the vendor partners, i.e. the integrators and consultants.

The unsolved part of the overall problem. The main feature of the approaches in use within the concept of Big Data is the ability to process the information array perfectly to obtain the most reliable results. Previously we had to rely on the so-called representative information retrieval or information subset. Consistently the inaccuracies in this approach were considerably higher. In addition, this approach required a certain amount of resource consumption in data preparation for the analysis and bringing them to the required format. It is recommended to develop a formal model of Big Data technology to simplify the modeling component of information systems based on their features.

## OBJECTIVES

The main objective of the paper is to develop a generalized formal model of the Big Data technology. It is necessary to formalize and describe the application features of the constituent elements of the model components.

## GENERALIZED FORMAL MODEL OF THE BIG DATA TECHNOLOGIES

Generalizing the listed definitions of the "Big data" term as an information technology, it is possible to develop a formal model in the form of a quartet, which looks like:

$$BD = \langle Vol_{BD}, Ip, A_{BD}, T_{BD} \rangle, \quad (1)$$

where: $Vol_{BD}$ is a set of volume types, $Ip$ is a set of types of data sources (information products), $A_{BD}$ is a set of techniques of Big Data analysis, $T_{BD}$ is a set of Big Data processing technologies.

Hinchcliff divides the approaches to the Big Data into three groups depending on the volume [3], which looks like:

$$Vol_{BD} = \{Vol_{FD}, Vol_{BA}, Vol_{DI}\}, \quad (2)$$

where: $Vol_{FD}$ is Fast Data – their volume is measured in terabytes, $Vol_{BA}$ is Big Analytics – they are petabyte data, $Vol_{DI}$ is Deep Insight – it is measured in exabytes, zettabytes.

Groups differ among themselves not only in the operated volumes of data, but also in the quality of their processing solutions. Fast Data processing does not provide new knowledge, its results are in line with the prior knowledge and enable to assess the process executions, to see better and in more detail what is happening, to confirm or reject some hypotheses. The velocity used for this technology must grow simultaneously with the growth of data volume.

Tasks solved by means of Big Analytics are used to convert the data recorded in the information into new knowledge. The system is based on the principle of the "supervised learning".

Processing information from different expressive power types of information sources, namely structured, semistructured, and unstructured is necessary for the Big Data technology. A set of information products is divided into three blocks, which looks like:

$$Ip = \langle St, SemS, UnS \rangle, \quad (3)$$

where: $St = \langle DB, DW \rangle$ is structured data (databases, warehouses), $SemS = \langle Wb, Tb \rangle$ is semi-structured data (XML, electronic worksheets), $UnS = \langle Nd \rangle$ is unstructured data (text).

There are operations and predicates on this vector and its separate elements that provide the transformation of various vector elements into each other, combining elements of the same type, search for items by the keyword. Computer programs are becoming closer to the real world in all its diversity, hence the growth in the volume of input data and hence the need for their analytics, moreover, in the mode maximally close to the real time. The convergence of these two trends has led to the emergence of an approach of Big Data Analytics.

Today, there is $A_{BD} = \{A_i\}$ set of different methods for the data set analysis, which are based on the tools borrowed from statistics and informatics (e.g. machine learning). The list is not exhaustive, but it reflects the most demanded approaches in various industries. Of course, the larger volume and diversified array are subject to analysis, the more accurate and relevant data are obtained at the output.

- Methods and techniques of the analysis applied to the Big Data are identified in the McKinsey report [9]:
- methods of Data Mining class: learning of association rules, classification (categorization methods of the new data on the basis of the principles previously applied to the existing data), cluster analysis, regression analysis,
- crowdsourcing is a categorization and data enrichment of the indefinite general public attracted on the basis of a public offer without entering into an employment relationship,
- mixing and integration of data is a set of techniques that allow us to integrate heterogeneous data from various sources for deeper analysis, for example, digital signal processing and natural language processing (including tonal analysis),
- machine learning, including supervised and unsupervised learning, as well as Ensemble learning that is the use of models based on the statistical analysis or machine learning to obtain a comprehensive prediction on the basis of the constituent models,
- artificial neural networks, network analysis, optimization, including genetic algorithms,
- pattern recognition,
- predictive analytics,
- simulation modeling,
- spatial analysis is a class of methods using topological, geometrical and geographical information in the data,
- statistical analysis, for example, A/B testing and time series analysis,
- visualization of analytical data is the presentation of information in the form of drawings, diagrams with the use of interactive features and animations of the results and for the use as the initial data for the subsequent analysis. Huge volumes combined with high velocity that distinguish Big Data Analytics from other programs require relevant computers and today virtually all major manufacturers offer specialized software and hardware systems.

New tools for analysis are necessary because there are not just more data than before but more of their external and internal sources, they are now more complex and diverse (structured, unstructured, and quasistructured), different indexing schemes are used (relational, multidimensional, NoSQL). It is impossible to cope with the data by the old man's, Big Data Analytics applies to large and complex arrays, so it is used the term of Discovery Analytics (which opens up Analytics) and Exploratory Analytics (which explains the analytics).

A collection of raw data uses the appropriate hardware and software technologies which depend on the control object nature (RFID, information from social networks, text documents, etc.) [1]. These data are transmitted to the input of the analytical machine. The controller is based on the hardware and software platform which has its own analytical software, it does not provide the generation of the controlling actions sufficient for the automatic control. DBMS analytics are primarily prognostic or predictive (Predictive Analysis, PA). Data accumulated in the data warehouses are output of the PA systems in most existing implementations. For the analysis the data are primarily moved to the Independent Data Mart (IDM) where: the data representation does not depend on the used applications. Then the same data are transferred to the specialized Analytical Data Mart (ADM) and they are processed with the specialists using various development tools or Data Mining. Such multistage model is acceptable for relatively small volumes of data, but when they increase with growing demand for the efficiency there are a number of shortcomings in these models. In addition to the need to move data, the presence of many independent ADM leads to the physical and logical infrastructure complication, a number of instruments used in the simulation increase, the results obtained by different analysts are inconsistent, computing power and channels are not optimally used. In addition, the separate existence of warehouses and ADM makes analytics almost impossible in close to real time.

The solution may be the approach called In-Database Analytics or No-Copy Analytics, which involves the use of data that are directly in the database. Such DBMS sometimes refer to as analytical and parallel. The approach is particularly attractive with the advent of MapReduce and Hadoop technology. In the new programme of the In-Database Analytics class generation all kinds of the data development and other types of intensive work are performed directly on the data in the warehouse. Obviously, this significantly speeds up the process and allows you to perform real-time applications such as pattern recognition, clustering, regression analysis, various kinds of forecasting. The acceleration is achieved not only by getting rid of the motion from the warehouse in a data mart, but mainly due to the use of different parallelization methods, including cluster systems with unlimited scalability. The solution such as In-Database Analytics opens the opportunity to use the cloud technologies in addition to analytics. The next step could be the technology of SAP HANA (High Performance Analytic Appliance) the essence of which is the data allocation in RAM.

On the phases of the Big Data processing the following technologies are used, which looks like:

$$T_{BD} = \langle T_{NoSQL}, T_{SQL}, T_{Hadoop}, T_V \rangle, \qquad (4)$$

where: $T_{NoSQL}$ is the technology of NoSQL databases, $T_{Hadoop}$ is the technology that ensures the massively-parallel processing, $T_{SQL}$ is the technology of the structured data processing (SQL database), $T_V$ is the technology of the Big Data visualization.

SN-architecture (Shared Nothing Architecture) is often stated as the basic principle of the Big Data processing that provides massively-parallel, scalable processing without degradation of hundreds or thousands of processing nodes [9, p. 31-33]. Thus, except for the technologies such as NoSQL, MapReduce, Hadoop, R considered by most analysts, McKinsey also considers Business Intelligence technologies and SQL supported relational database management system in terms of Big Data applicability.

NoSQL (Not only SQL) in computer science is a number of approaches aimed at implementing database warehouses which differ from the models used in the traditional relational DBMS with access to the data by SQL means. It applies to the databases in which an attempt is made to solve the problems of scalability and availability at the expense of data atomicity and consistency.

MapReduce is an engine of parallel processing. MapReduce is a model of distributed computing introduced by Google, which is used for parallel computing on very large data sets in computer clusters. MapReduce is a computing framework for the distributed task sets using a large number of computers that make up the cluster.

Hadoop is a project of the Apache Software Foundation, the freely available set of utilities, libraries and frameworks for the development and execution of the distributed programs running on the clusters of hundreds or thousand nodes. It is developed in Java within the computing MapReduce paradigm according to which the application is divided into a large number of identical elementary tasks that are feasible on the cluster nodes and naturally given in the final result. Hadoop is a capability set with open source code. Hadoop handles the large data caches breaking them into smaller, more accessible to applications and distribution across multiple servers for the analysis. Hadoop processes requests and generates the requested results in significantly less time than the old school of software analytics, it often takes minutes instead of hours or days.

As of 2014, the project consists of four modules: Hadoop Common (software for integration, a set of infrastructure software libraries and utilities used for other modules and related projects), HDFS (distributed file system), YARN (a system for task scheduling and cluster management) and Hadoop MapReduce (a programming and execution platform of the distributed MapReduce computing).

One of the technologies that should be used for Big Data is the data space. The data space is a block vector containing a variety of information products of the subject domain [10, 11].

A visual representation of the Big Data analysis results is crucial to their interpretation. There is no secret that human perception is limited, and scientists continue to conduct research in the field of improvement of the modern methods of data presentation in the form of images, diagrams or animations. We describe a few of the latest innovative visualization techniques.

The tag cloud. Each element in the tag cloud is assigned a specific weight factor which correlates with the font size. In the case of text analysis the weight factor value depends on the use (citation) frequency of a particular word or phrase. It allows the reader to quickly get an idea about the key points of the arbitrarily large text or set of texts.

*Clustergram*. It is the visualization method used for cluster analysis. It shows how individual elements in the data set correspond to the clusters as their quantity change. The choice of the optimal cluster number is an important component of the cluster analysis.

*Historical flow*. It helps to monitor the document evolution. A large number of authors simultaneously create this document. In particular, this is a typical situation for the wiki services. The horizontal axis indicates time, the vertical one indicates the contribution of each of the co-authors, i.e. the volume of the entered text. Each unique author is assigned a specific color on the chart.

*Spatial information flow*. This chart allows you to track the spatial information distribution.

Unprecedented variety of data resulting from the huge number of possible transactions and interactions provide an excellent foundation for business on forecasts amendments, assessment of the development prospects of products and entire approaches, better cost control, performance assessment, and the like. On the other hand, the Big Data pose the challenging tasks for any IT department.

IT manager who intends to benefit from large structured and unstructured data must comply with the following technical considerations [5, 12, 13]:

- divide and conquer. Data transfer and integration are necessary, but both approaches increase capital and operating costs for the information extraction tools, its transformation and loading (ETL), so, do not neglect standard relational environments and analytical data warehouses,
- compression and deduplication. You should always remember what part of the compressed data may require recovery, and starting out from each specific situation you make a decision about using the same compression,
- not all data are equal. Depending on the specific situation the range of queries for the business analytics varies within wide limits. Often to obtain the necessary information is sufficient to answer the SQL query, but there are also deep analytical queries that require the business intelligence tools and have the all features of the dashboard and visualization. To prevent a sharp increase in operating expenses you need to be careful about the compilation of a balanced list of the necessary patented technologies in combination with the open-source Apache Hadoop,
- scalability and manageability.

Organizations have to solve the problem of heterogeneity of databases and analytical environments, and therefore the ability to scale horizontally and vertically is of fundamental importance. The easiness of the horizontal scaling became one of the main reasons for the rapid Hadoop distribution.

Craig Bati, the executive marketing director and director in Fujitsu Australia technology, indicates that the business analysis is a descriptive analysis of the results achieved by the business in a certain period of time, while the velocity of Big Data processing allows you to analyze the root to offer the business some recommendations for the future. Big Data technologies also allow analyzing more data types in comparison with business analytics tools that give you the opportunity to focus not only on the structured data warehouses.

Big Data are designed to handle the greater amounts of information than business analytics and it certainly fits the traditional definition of big data. Big Data are designed for processing of more quickly receiving and changing information; it means a deep exploration and interactivity. In some cases results are generated faster than a web page loading. Big Data are designed to handle the unstructured data which use we only begin to learn after they could establish their collection and storage, and we need the algorithms and the possibility of dialogue to facilitate the trends search inside these arrays. Big Data handling is not similar to the normal process of business analytics where: the simple addition of known values gives the result. When working with the Big Data we obtain the result in the process of their cleaning by sequential simulation: first, the hypothesis are offered, then statistical, visual or semantic model is developed; with reference to this model the offered hypothesis is tested and then the next one is offered. This process requires from the researcher either the interpretation of visual values or making interactive queries on the basis of knowledge, or the development of adaptive "machine learning" algorithms that are able to get the desired result. Moreover, the lifetime of this algorithm can be quite short.

By ignoring the role of information and data as a subject of research the same mine was planted that exploded now, at a time when the needs changed, when the computational load on the computers was much less than other types of works to be performed on the data. The goal is to obtain a new information and new knowledge from existing data sets. That's why it is meaningless to talk about solving the problem of Big Data within the restoration of the "information - data - knowledge" chain links. The information is processed to obtain the data that should be just enough so that a man could transform them into knowledge.

CONCLUSIONS

So, Big Data are not a speculation, but a symbol of the coming technological revolution. The need for the analytical effort with Big Data will significantly change the face of the IT industry and stimulate the emergence of the new software and hardware platforms. To achieve the desired goal, the development of a formal model of the Big Data information technology is made and its structural elements are described. Today for the analysis of large volumes of data the most advanced methods are used: artificial neural network (models are built on the principle of the biological neural network organization and functioning); predictive analytics, statistics and Natural Language Processing (areas of artificial intelligence and mathematical linguistics that study the problems of computer analysis and natural language synthesis). Also, we use the methods that engage human experts, crowdsourcing, $A/B$ testing, sentiment analysis, and the like. To visualize the results the known methods are applied, for example, tag clouds and completely new Clustergram, History Flow and Spatial Information Flow. Big Data is supported by the distributed file system Google File System technology, Cassandra, HBase, Lustre and ZFS, by the MapReduce and Hadoop programming constructs and many other solutions. According to the experts, such as McKinsey Institute, the manufacturing, healthcare, trade, administration and control of individual movements undergo the transformations under the influence of the Big Data. Further research will be devoted to the study of methods, models and tools to effectively support the overall activity of the model development.